\documentclass[a4paper,preprint]{emulateapj}
\usepackage{amstext,amsmath,amssym}

\shorttitle{Build--up of the red--sequence since z$\sim 0.8$}
\shortauthors{De Lucia et al.}

\begin{document}


\title{The build--up of the red--sequence in galaxy clusters since $z\sim
  0.8$\footnote{\small Based on observations obtained at the ESO Very Large
    Telescope (VLT) as part of the Large Programme 166.A--0162 (the ESO Distant
    Cluster Survey).}}


\author{G. De Lucia\altaffilmark{1}, B. M. Poggianti\altaffilmark{2}, A.
  Arag\'on-Salamanca\altaffilmark{3}, D. Clowe\altaffilmark{4}, C.
  Halliday\altaffilmark{2}, P. Jablonka\altaffilmark{5}, B.
  Milvang-Jensen\altaffilmark{6}, R. Pell\'o\altaffilmark{7}, S.
  Poirier\altaffilmark{5}, G. Rudnick\altaffilmark{1}, R.
  Saglia\altaffilmark{6}, L.  Simard\altaffilmark{8}, S.~D.~M.
  White\altaffilmark{1}}

\altaffiltext{1}{Max--Planck--Institut f\"ur Astrophysik,
  Karl--Schwarzschild--Str. 1, 85748 Garching bei M\"unchen, Germany}
\altaffiltext{2}{Osservatorio Astronomico di Padova, vicolo dell'Osservatorio
  5, 35122 Padova, Italy} 
\altaffiltext{3}{School of Physics and Astronomy, University of Nottingham, NG7
  2RD,UK} 
\altaffiltext{4}{Steward Observatory, University of Arizona, 933 North Cherry
  Avenue, Tucson, AZ 85721} 
\altaffiltext{5}{Observatoire de Paris, 5 place Jules Janssen, F-92195 Meudon
  Cedex, France} 
\altaffiltext{6}{Max-Planck-Institut f\"ur Extraterrestrische Physik,
  Giessenbachstr., 85748 Garching, Germany} 
\altaffiltext{7}{Laboratoire d'Astrophysique, UMR 5572, Observatoire
  Midi-Pyr\'en\'ees, 14 Avenue E. Belin, 31400 Toulouse, France}
\altaffiltext{8}{Herzberg Institute of Astrophysics, National Research Council
  of Canada, Victoria, BC V9E 2E7, Canada}

\begin{abstract}
  We study the rest--frame (U$-$V) color--magnitude relation in $4$ clusters at
  redshifts $0.7$--$0.8$, drawn from the ESO Distant Cluster Survey.  We
  confirm that red--sequence galaxies in these clusters can be described as an
  old, passively--evolving population and we demonstrate, by comparison with
  the Coma cluster, that there has been significant evolution in the stellar
  mass distribution of red--sequence galaxies since $z\sim 0.75$. The EDisCS
  clusters exhibit a deficiency of low luminosity passive red galaxies.
  Defining as `faint' all galaxies in the passive evolution corrected range
  $0.4\,\gtrsim\,$L$/$L$_*\,\gtrsim\,0.1$, the luminous--to--faint ratio of
  red--sequence galaxies varies from $0.34\pm0.06$ for the Coma cluster to
  $0.81\pm0.18$ for the high redshift clusters.  These results exclude a
  synchronous formation of all red--sequence galaxies and suggest that a large
  fraction of the faint red galaxies in current clusters moved on to the red
  sequence relatively recently.  Their star formation activity presumably came
  to an end at $z\lesssim0.8$.
\end{abstract}





\section{Introduction}
\label{sec:intro}

It has long been known \citep{vs} that red cluster galaxies form a tight
sequence in the color--magnitude diagram that, in nearby clusters, extends at
least $5$--$6$ mag fainter than the Brightest Cluster Galaxy (BCG)
\citep{depropris}.  The existence of a tight color--magnitude relation (CMR)
for cluster early type galaxies up to redshift $\sim 1$, and the evolution of
its slope and its zero--point as a function of redshift, are commonly
interpreted as the result of a formation scenario in which cluster elliptical
galaxies constitute a passively evolving population formed at high redshift
($z\gtrsim2$--$3$) \citep*{ellis,stanford,gladders}.  In this model, the slope
of the relation reflects metallicity differences and naturally arises through
the effects of supernova winds \citep{kodama}. An alternative explanation has
been proposed by \citet[][ see also \citet{me2}]{kauffcharl}.  In this model,
elliptical galaxies form through mergers of disk systems and a CMR arises as a
result of the fact that more massive ellipticals are formed by mergers of more
massive, and hence more metal rich, disk systems.

The homogeneity of cluster elliptical galaxies should, however, be contrasted
with the increasing observational evidence that red passive galaxies in distant
clusters constitute only a subset of the passive galaxy population in clusters
today \citep{df}.  Distant clusters contain significant populations of galaxies
with active star formation, that must later have evolved on to the CM sequence
after their star formation activity was terminated, possibly as a consequence
of their environment \citep{smail,dressler99,poggianti99}.

In this Letter we present the CMR for $4$ clusters in the redshift interval
$0.7$--$0.8$ from the ESO Distant Cluster Survey (hereafter EDisCS).  In the
following, we use: $\Omega_{\rm m} = 0.3$, $\Omega_{\Lambda}=0.7$ and $H_0 =
70\,{\rm km}\,{\rm s}^{-1}\,{\rm Mpc}^{-1}$.
\section{The data}
\label{sec:observations}

EDisCS is an ESO Large Programme aimed at the study of cluster structure and
cluster galaxy evolution over a significant fraction of cosmic time.  The
complete EDisCS dataset provides homogeneous photometry and spectroscopy
for $10$ clusters at redshift $0.4\div0.5$ and $10$ clusters at redshift
$0.6\div0.8$.  In this Letter we present results for $4$ clusters at redshift
$0.7$ (cl$1040.7$--$1155$ and cl$1054.4$--$1146$), $0.75$ (cl$1054.7$--$1245$),
and $0.8$ (cl$1216.8$--$1201$).

The clusters used in this analysis have been imaged in V, R, and I with FORS2
($6\farcm8\times6\farcm8$) on the ESO Very Large Telescope (VLT).  The average
integration times used were $115$ m in the I--band, and $120$ m in the V--band.
Multi--object spectroscopy was carried out using FORS2 on VLT.  The clusters in
the high redshift sample (that includes those used in this study) have also
been imaged in J and K (SOFI-NTT, $5\farcm5\times5\farcm5$).  A brief
description of the cluster sample selection is given in \citet{rudnick03}.  All
data and details about their analysis will be presented in forthcoming papers
(White et al, Halliday et al., Arag\'on-Salamanca et al., in preparation).  The
number of spectroscopically confirmed members for the clusters presented in
this analysis ranges from $30$ (cl$1040.7$--$1155$) to $70$
(cl$1216.9$--$1136$).

Object catalogs have been created using the SExtractor software \citep{bertin}
in `two--image' mode using the I--band images as detection reference images.
In the following, we will use magnitude and colors measured on the
seeing--matched (to $0\farcs8$ -- our `worst' seeing) registered frames using a
fixed circular aperture with $1\farcs0$ radius.  At the clusters redshifts,
this corresponds to a physical radius of $7.13$--$7.50$ kpc.  This choice has
been adopted in order to simplify the comparison with the Coma cluster (see
Sec.~\ref{sec:cm}).

\section{Cluster membership}
\label{sec:photoz}

Photometric redshifts were computed using two independent codes
\citep*{greg,roser} to better keep under control systematics in the
identification of likely non--members.  A detailed analysis of the performance
of both codes on EDisCS data will be presented in a separate paper (Pell\'o et
al., in preparation).  Results based on the current spectroscopic sample show
that our photometric redshifts are quite accurate, with $<|z_{\rm spec}-z_{\rm
  phot}|> =0.06$--$0.08$.  Unless otherwise stated, non--members are rejected
in the following using a two-step procedure: (i) the full redshift probability
distributions from both codes are used to reject objects with low probability
to be at the cluster redshift \citep[e.g.][]{brunner}; (ii) a statistical
subtraction is performed on the remaining objects using the distribution on the
CM diagram of all the objects at a physical distance from the BCG larger than
$1$ Mpc \citep{kodbow}. Our photometric redshift selection rejects $85-90\%$ of
the objects in the original photometric catalogs (containing $2600-3100$
objects, down to a magnitude limit of $25$ in the I--band).  $40-50\%$ of the
remaining objects are subtracted statistically.

\section{The color--magnitude relation}
\label{sec:cm}

\begin{figure}
\epsscale{1.2}
\plotone{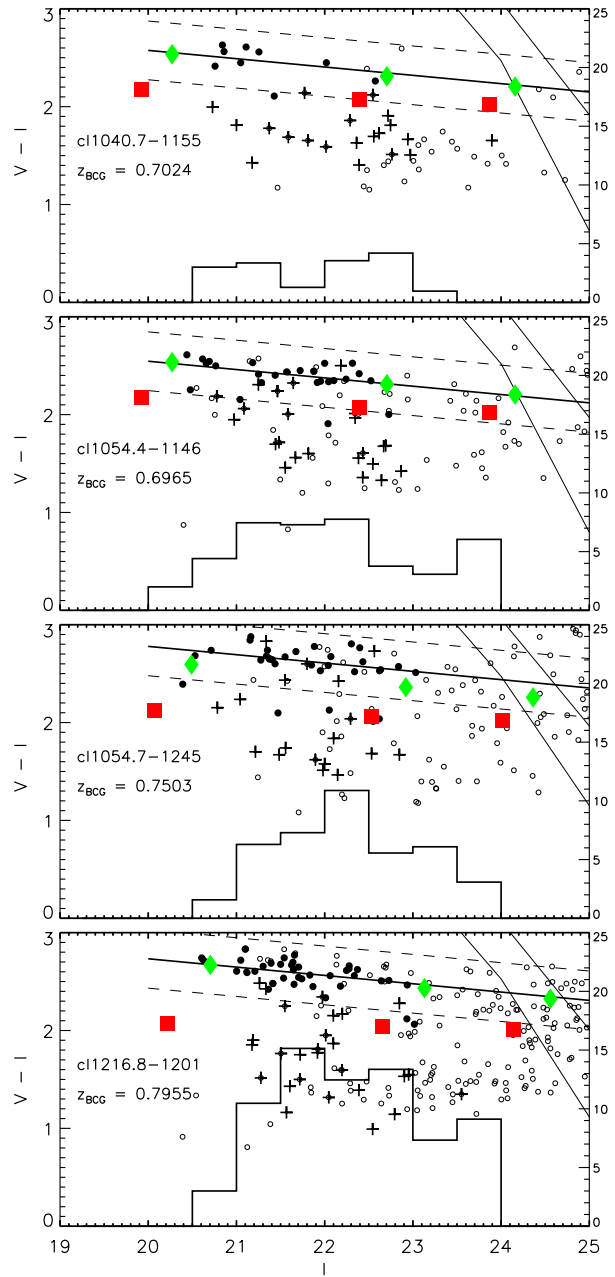}
\caption{CM diagrams for the $4$ clusters under investigation.  Solid thick
  line represents the best fit relations measured using the bi--weight
  estimator \citep*{beers} and a fixed slope of $-0.085$.  Dashed lines
  correspond to $\pm 3\sigma$ from the best fit line, where $\sigma$ is the
  dispersion of the objects used for the fit ($\sim 0.1$ in all $4$ clusters).
  Thin solid lines correspond to the $3$ and $5\sigma$ detection limits in
  V$-$I color.  Empty symbols represent galaxies retained as cluster members
  after using their photometric redshifts and a particular Monte Carlo
  realization of the statistical subtraction.  Filled circles are
  spectroscopically confirmed members with absorption--line spectra, while
  crosses represent spectroscopically confirmed members with emission--line
  spectra.  Filled squares and diamonds represent two families of models (see
  text for details).  Spectroscopic redshifts are used in membership
  determination, where available. The solid histogram in each panel represents
  the number of objects within $3\sigma$ from the best--fit red--sequence after
  averaging over $100$ Monte Carlo realizations of the statistical subtraction.
  \label{cm}}
\end{figure}

Fig.~\ref{cm} shows the V$-$I diagram for the $4$ clusters used in this
analysis.  At the clusters redshifts, V$-$I approximately samples the
rest--frame U$-$V color, which straddles the $4000$\AA~Balmer break, therefore
being very sensitive to any recent or ongoing star formation (SF).  A red
sequence is clearly visible in the CM diagram of each cluster, together with a
significant population of the blue galaxies known to populate high redshifts
clusters.

\begin{figure}
\epsscale{1.2}
\plotone{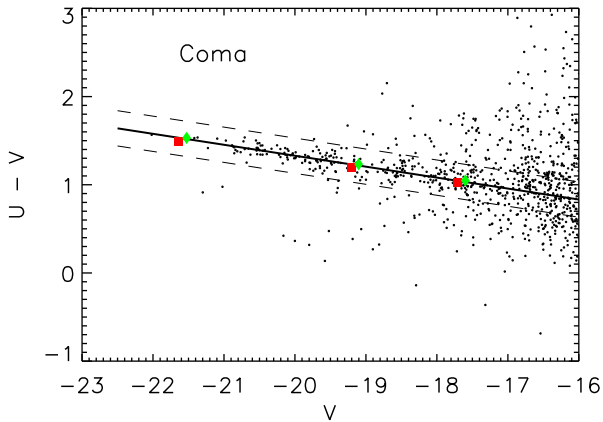}
\caption{The CMR of Coma for all the galaxies in the catalog by \citet{terl}.
  Observed magnitudes are converted to absolute magnitudes using the
  distance modulus of Coma ($35.16$) and observed colors are converted to
  rest--frame colors using tabulated K--corrections \citep{poggianti97}.
  Solid, dashed lines, and symbols have the same meaning as in
  Fig~\ref{cm}.\label{coma}}
\end{figure}

Filled squares and diamonds in Fig.~\ref{cm} show the location of galaxy models
with two different SF histories: a single burst at $z=3$ (diamonds) and an
exponentially declining SF starting at $z=3$ (squares) with $\tau=1$ Gyr.
Both were calculated with the population synthesis code by \citet{BC}.  For
each SF history, three different metallicities are shown: $0.02$, $0.008$ and
$0.004$, going from brighter to fainter objects.  The relation between
metallicity and luminosity in these models has been calibrated by requiring
that they reproduce the observed CMR in Coma\footnote{This calibration has been
  found {\it a posteriori} to be in good agreement with the
  metallicity--luminosity relation derived from spectral indices of Coma
  galaxies (Poggianti et al.  2001).}, as shown in Fig.~\ref{coma}.  Here we
use the data from \citet*{terl} and use their magnitude and colors in a
$25\farcs2$ diameter aperture.  At the redshift of Coma, this corresponds to a
physical size of $11.71$ kpc, quite closely approximating our $\sim 14$ kpc
aperture at $z\sim 0.8$.  Fig.~\ref{cm} shows that the single burst model
provides a remarkably good fit to the red sequence observed in the high
redshift clusters, confirming that the location of the CM sequence observed in
distant clusters requires high redshifts of formation, and that the slope is
consistent with a correlation between galaxy metal content and luminosity.

\section{The truncation of the red sequence}
\label{sec:compare}

Perhaps the most interesting result of our analysis is that the red
sequence in our clusters is well populated at magnitudes brighter than $\sim
22$, but unusually `empty' at fainter magnitudes.  The histograms in
Fig.~\ref{cm} show the number of cluster members within $\sim 3\sigma$ from the
best fit relation, after averaging over $100$ Monte Carlo realizations of the
statistical subtraction.  The paucity of low--luminosity red galaxies at
high--z occurs at magnitudes well above our completeness limit and in all the
clusters under investigation.  There is no clear evidence for such a deficiency
in cl$1040.7$--$1155$ ($z=0.7$) but, given its very low fraction of objects
with absorption--line spectra, it cannot be ruled out statistically either.

\begin{figure}
\epsscale{1.2}
\plotone{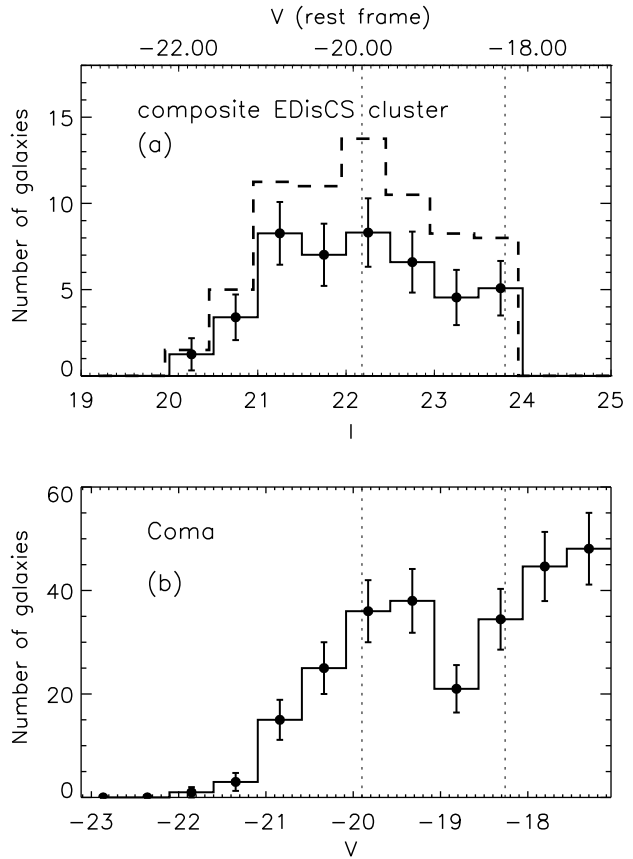}
\caption{Average number of galaxies on the red-sequence as a function of
  magnitude for our composite cluster at redshift $0.75$ (panel a) and the Coma
  cluster (panel b).  Error bars represent Poisson errors.  The scale on the
  top of panel (a) shows the rest--frame V--band magnitude corresponding to the
  observed I--band magnitudes after correcting for passive evolution between
  $z=0.75$ and $z=0$. The dashed histogram in panel a is obtained when
  membership is based solely on photometric redshifts and no statistical
  correction is applied. A small shift in the x--coordinate has been applied to
  make the plot more easily readable. 
  \label{histo}}
\end{figure}

In order to quantify this effect, we combine the histograms shown in
Fig.~\ref{cm} correcting colors and magnitudes to a common redshift $=0.75$.
The result is shown in panel (a) of Fig.~\ref{histo}.  Panel~(b) shows the
corresponding result for the Coma cluster. For the high--z clusters,
non--members have been excluded as described in \S3. The dashed histograms
shows the corresponding result when membership is based solely on photometric
redshifts. In Coma, membership information is available from spectroscopy for a
large number of galaxies and we have corrected the number of red galaxies in
each magnitude bin for background and foreground contamination using a redshift
catalog kindly provided by Matthew Colless and the same procedure as in
\citet{mobasher}.

The histogram of the EDisCS clusters shows a decrease in the number of
$\lesssim 0.4$L$_*$ red--sequence galaxies.  This effect is present in all our
clusters individually, with the possible exception of cl$1040.7$--$1155$, as
discussed above.  Indeed, the `luminosity function' of the red galaxies in
these clusters shows the same decrease despite the variety in cluster
properties such as velocity dispersion, richness, concentration and
substructure (White et al.; Halliday et al., in preparation).  Unfortunately,
such a deficiency coincides with the well--known `dip' in the luminosity
function of the Coma cluster \citep{GP}.  However, the behaviour of the Coma
cluster seems quite `untypical'. \citet*{depropris03}, for example, have shown
that the luminosity function of early (passive) spectral types increases going
to fainter magnitudes in clusters in the 2dF survey. 

After correcting for passive evolution with the single burst model shown in
Figs.~\ref{cm} and \ref{coma}, the histograms shown in Fig.~\ref{histo} are
significantly different.  A Kolmogorov--Smirnov test rejects the hypothesis
that the two histograms are drawn from the same parent distribution at the
$\sim 95\%$ level.  Our correction for passive evolution is $\sim 0.8$ mag at V
from $z=0$ to $z=0.75$, and is consistent with studies of the Fundamental Plane
\citep{vd}.  An increase in this correction by $0.2$ mag or more lowers the
significance of the effect to $\sim 80\%$.

If we arbitrarily classify as `luminous' galaxies those brighter than $M_{\rm
  V}= -19.9$ (this corresponds to an observed I--band magnitude of $22.18$ at
redshift $0.75$), and as `faint' those galaxies that are fainter than this
magnitude and brighter than $-18.36$ (this corresponds to the limiting
magnitude in the I--band for which all the selected objects are above the
$5\sigma$ detection limit in the V--band), we obtain a luminous--to--faint
ratio for the Coma cluster $0.34 \pm 0.06$ (the error has been estimated
assuming Poisson statistics and, for the EDisCS clusters, includes the error
associated to the statistical field subtraction).  The corresponding value of
the luminous--to--faint ratio for the composite EDisCS cluster is equal to
$0.81\pm0.18$.  This ratio is therefore different between the composite high
redshift cluster and the Coma cluster at about the $3\sigma$ level.

The results are robust against both the technique adopted for removing
non--cluster members and the photometric errors.  The red galaxy deficit is
detected also when rejecting non--members using a purely statistical
subtraction or a more stringent criterion for membership based solely on
photometric redshifts.  In fact, a deficit is evident also in the {\it full}
photometric catalog, when no field correction is attempted.  Photometric errors
in the EDisCS catalog are comparable to the errors in Terlevich data, therefore
the differences observed in the distributions of Fig.~\ref{histo} cannot be a
spurious result arising from photometric errors.  As a further test, we
performed a series of Monte Carlo realizations ($100$) where each point in the
Coma histogram was scattered around the CMR taking into account the typical
error on color and magnitude. The shape of the histogram is not found to vary
significantly.

\section{Discussion}
\label{sec:discussion}
A decline in the number of red sequence members at faint magnitudes was first
observed in clusters at $z=0.25$ by \citet{smail}.  Evidence for a `truncation'
of the red--sequence has been noticed in a cluster at $z=1.2$ by \citet{kaj}
and \citet{nakata}.  There have, however, been speculations from the same
authors that such a result was spurious because of limited area coverage
($<0.33$ Mpc) and strong luminosity segregation.  (See the discussion by
\citet{K04} who obtain a similar result for early--type galaxies in a single
deep field).

The CMRs of the EDisCS clusters at $z\sim0.8$ show a deficiency of red,
relatively faint galaxies, and suggest that such a deficit may be a universal
phenomenon in clusters at these redshifts.  Our investigation shows that a
large fraction of present--day passive $\lesssim 0.4$L$_*$ galaxies must have
moved on to the CMR at redshifts lower than $0.8$.  Their SF activity therefore
must have ended after $z\sim 0.8$.  The physical mechanisms and the
characteristic time--scales of this transformation are not yet understood.  The
populations of blue galaxies observed in distant clusters are the logical
progenitors for a significant fraction of the faint red galaxies at $z=0$
\citep{smail,kodbow,deprop}.

It is clear that a formation scenario in which all red galaxies in clusters
evolved passively after a synchronous monolithic collapse at $z\gtrsim 2$--$3$,
is inconsistent with observations.  Our results indicate that present day
passive galaxies follow different evolutionary paths, depending on their
luminosity.  Future studies, including the intermediate redshift clusters in
the EDisCS sample, will help to understand the relative importance of star
formation and metallicity in establishing the observed red--sequence.




\acknowledgments We thank R. Bender, S. Charlot, M. Colless, G. Kauffmann, F.
La Barbera, M. Pannella, I. Smail, V. Strazzullo, S. Zaroubi, D. Zaritsky and
S. Zibetti.  G.~D.~L.  acknowledges financial support from the Alexander von
Humboldt Foundation, the Federal Ministry of Education and Research, and the
Programme for Investment in the Future (ZIP) of the German Government and the
hospitality of the Osservatorio Astronomico di Padova.




\begin{thebibliography}{}
\harvarditem[Beers et al.]{Beers, Flynn, \& Gebhardt}{1990}{beers} 
  Beers, T. C., Flynn, K.,  Gebhardt, K. 1990, AJ, 100, 32
\bibitem[\protect\citename{Bertin \& Arnouts, }1996]{bertin}
  Bertin, E., Arnouts S. 1996, A\&AS, 117, 393
\harvarditem[Bolzonella et al.]{Bolzonella, Miralles, \& Pell\'o}{2000}{roser}
  Bolzonella, M., Miralles, J.-M., Pell\'o, R. 2000, A\&A, 363, 476
\bibitem[\protect\citename{Brunner \& Lubin, }2000]{brunner}
  Brunner, R. J., Lubin, L. M. 2000, ApJ, 120, 2851
\bibitem[\protect\citename{Bruzual \& Charlot, }2003]{BC}
  Bruzual, G., Charlot, S. 2003, MNRAS, 344, 1000
\harvarditem[De Lucia et al.]{De Lucia, Kauffmann, \& White}{2004}{me2} 
  De Lucia, G., Kauffmann, G., White, S. D. M. 2004, MNRAS, 349, 1101 
\bibitem[\protect\citename{De Propris et al., }1998]{depropris}
  De Propris, R., Eisenhardt, P. R., Stanford, S. A., Dickinson, M. 1998, ApJ,
  503, L45 
\harvarditem[De Propris et al.]{De Propris, Colless, \&
  Driver}{2003}{depropris03}   
  De Propris, R., Colless, M., Driver, S.P. 2003, MNRAS, 342, 725
\bibitem[\protect\citename{De Propris et al., }2003]{deprop} 
  De Propris, R., Stanford, S. A., Eisenhardt, P. R., Dickinson, M. 2003, ApJ,
  598, 20
\bibitem[\protect\citename{van Dokkum and Franx, }1996]{df}
  van Dokkum, P. G., Franx, M. 1996, MNRAS, 281, 985
\bibitem[\protect\citename{Dressler et al., }1999]{dressler99}
  Dressler, A., Smail, I., Poggianti, B. M., Butcher, H., Couch, W. J., Ellis,
  R. S, Oemler, A. Jr 1999, ApJS, 122, 51
\bibitem[\protect\citename{Ellis et al., }1997]{ellis}
  Ellis, R. S., Smail, I., Dressler, A., Couch, W. J., Oemler, A. Jr, Butcher,
  H., Sharples, R. M. 1997, ApJ, 483, 582
\bibitem[\protect\citename{Gladders et al., }1998]{gladders}
  Gladders, M. D., L\'opez-Cruz, O., Yee, H. K. C., Kodama, T. 1998, ApJ, 501,
  571 
\bibitem[\protect\citename{Godwin \& Peach, }1977]{GP}
  Godwin, J. G., Peach, J. V., 1977, MNRAS, 181, 323
\bibitem[\protect\citename{Kajisawa et al., }2000]{kaj}
  Kajisawa, M., Yamada, T., Tanaka, I., Maihara, T., Iwamuro, F., Terada, H.,
  Goto, M., Motohara, K., et al. 2000, PASJ, 52, 61
\bibitem[\protect\citename{Kauffmann \& Charlot, }1998]{kauffcharl}
  Kauffmann, G., Charlot, S. 1998, MNRAS, 294, 705
\bibitem[\protect\citename{Kodama et al., }1998]{kodama}
  Kodama, T., Arimoto, N., Barger, A. J., Arag\'on-Salamanca, A. 1998, A\&A,
  334, 99 
\bibitem[\protect\citename{Kodama \& Bower, }2001]{kodbow}
  Kodama, T., Bower, R. G. 2001, MNRAS, 321, 18
\bibitem[\protect\citename{Kodama et al., }2004]{K04}
  Kodama, T., Yamada, T., Akiyama, M., Aoki, K., Doi, M., Furusawa, H., Fuse,
  T., Imanishi, M., et al. 2004, MNRAR, in press, preprint astro-ph/0402276
\bibitem[\protect\citename{Mobasher et al., }2003]{mobasher}
  Mobasher, B., Colless, M., Carter, D., Poggianti, B. M., Bridges, T. J.,
  Kranz, K., Komiyama, Y., Kashikawa, N.,  et al. 2003, ApJ, 587, 605
\bibitem[\protect\citename{Nakata et al., }2001]{nakata}
  Nakata, F., Kajisawa, M., Yamada, T., Kodama, T., Shimasaku, K., Tanaka, I.,
  Doi, M., Furusawa, H., et al. 2001, PASJ, 53, 1139
\bibitem[\protect\citename{Poggianti, }1997]{poggianti97}
  Poggianti, B. M. 1997, A\&AS, 122, 399
\bibitem[\protect\citename{Poggianti et al., }1999]{poggianti99}
  Poggianti, B. M., Smail, I., Dressler, A., Couch, W. J., Barger, A. J.,
  Butcher, H., Ellis, R. S., Oemler, A. Jr. 1999, ApJ, 518, 576
\bibitem[\protect\citename{Rudnick et al., }2001]{greg}
  Rudnick, G., Franx, M., Rix, H., Moorwood, A., Kuijken, K., van Starkenburg,
  L., van der Werf, P., R\"ottgering, H., et al. 2001, ApJ, 122, 2205 
\bibitem[\protect\citename{Rudnick et al., }2003]{rudnick03} 	
  Rudnick, G., White, S., Arag\'on-Salamanca, A., Bender, R., Best, P., Bremer,
  M., Charlot, S., Clowe, D., et al. 2003, The Messenger, 112, 19
\bibitem[\protect\citename{Smail et al., }1998]{smail}
  Smail, I., Edge, A. C., Ellis, R. S., Blandford, R. D. 1998, MNRAS, 293, 124
\harvarditem[Stanford et al.]{Stanford, Eisenhardt, \&
  Dickinson}{1998}{stanford}   
  Stanford, S. A., Eisenhardt, P. R., Dickinson, M. 1998, ApJ, 492, 461
\harvarditem[Terlevich et al.]{Terlevich, Caldwell, \& Bower}{2001}{terl} 
  Terlevich, A. I., Caldwell, N., Bower, R. G. 2001, MNRAS, 326, 1547
\bibitem[\protect\citename{Visvanathan \& Sandage, }1977]{vs}
  Visvanathan, N., Sandage, A. 1977, ApJ, 216, 214
\bibitem[\protect\citename{Wuyts et al., }2004]{vd}
  Wuyts, S., van Dokkum, P. G., Kelson, D. D., Franx, M., Illingworth,
  G. D. 2004, ApJ, 605, 677 
\end{thebibliography}
\end{document}